\documentstyle[fleqn,12pt]{article}

\def\pj{\hspace{-.26cm}}
\def\fpj{\hspace{-.7cm}}
\def\thalf{{\textstyle{\frac{1}{2}}}}

\def\twothird{{\textstyle{\frac{2}{3}}}}
\def\tquar{{\textstyle{\frac{1}{4}}}}
\def\ttquar{{\textstyle{\frac{3}{4}}}}

\def\qsl{q\hspace{-2mm}/}
\newcommand{\vm}[1]{\mbox{\bf#1}}
\newcommand{\vmg}[1]{\mbox{\boldmath$#1$}}
\begin{document}
\title{An Effective Lagrangian with Broken Scale and Chiral Symmetry II:
\protect\\  Pion Phenomenology}
\author{G. Carter, P.J. Ellis and S. Rudaz \\[2mm]
{\small School of Physics and Astronomy}\\[-2mm] {\small University of
Minnesota, Minneapolis, MN 55455}}
\date{~}
\maketitle
\centerline{{\small\bf Abstract}}
\vskip-.15cm
{\small We extend an effective Lagrangian embodying broken scale and
chiral symmetry to include explicit chiral symmetry breaking and an
additional chiral invariant term which allows for an axial coupling constant
greater than unity. We also include a chiral Lagrangian for the isotriplet
vector mesons which leads to a renormalization of the pion field. The
properties of nuclear matter and nuclei, low energy $\pi N$ scattering and the
behavior of quantities such as the pion mass and axial coupling at finite
density are discussed. }
\thispagestyle{empty}
\vskip-19cm
\hfill NUC-MINN-95/21-T\\
\phantom{111}\hfill UMN-TH-1418/95\\
\phantom{111}\hfill December 1995\\
\newpage
\section{Introduction}

Since spontaneously broken chiral symmetry is thought to be a fundamental
feature of low energy effective Lagrangians, it is remarkable that it is
difficult to describe nuclear matter and nuclei with such a Lagrangian.
For example, it has long been known \cite{kerman} that the standard
Lagrangian of the
linear sigma model, supplemented by the repulsion of the $\omega$ meson, does
not yield a normal saturating equation of state for nuclear matter. In
order to describe nuclear matter, the Lagrangian can be modified by
generating the vector meson masses through coupling to the scalar $\sigma$
field or by including one-loop vacuum corrections. However, it is found
\cite{glu3,fs2,fs3} that such models fail to reproduce the observed
properties of nuclei.

In a previous paper \cite{glu3}, hereinafter referred to as I, we were
able to satisfactorily describe nuclear matter and finite nuclei with
an effective Lagrangian which incorporated broken scale symmetry in
addition to spontaneously broken chiral symmetry, as suggested by quantum
chromodynamics (QCD). In order to obtain good phenomenology it was necessary to
generate the vector meson masses by coupling to the glueball field, $\phi$, and
to discard the conventional ``Mexican hat" potential of the linear sigma model,
$\tquar\lambda\left(\sigma^2+\vmg{\pi}^2-f_{\pi}^2\right)^{\!2}$.
In fact by introducing an $(\omega_{\mu}\omega^{\mu})^2$ term in the Lagrangian
and adjusting the coupling, we obtained results of a quality comparable to
those of the standard Walecka model with non-linear $\sigma^3$ and
$\sigma^4$ terms \cite{fern2}. In our model the scalar masses are 1.5 GeV
and approximately 0.5 GeV
for the states which are predominantly $\phi$ and $\sigma$, respectively.
Here the $\sigma$ is the chiral partner of the pion and also provides
intermediate range attraction. This view of the $\sigma$ meson finds support
in the work of the Brooklyn
group \cite{brook} using the Nambu-Jona-Lasinio model to study correlated
two-pion exchange. They find that for the spacelike momentum transfers
of interest here, the $T$ matrix can be represented by an effective low-mass
sigma meson which is the chiral partner of the pion, while in the timelike
region a physical, low-mass scalar particle is not obtained. On the other
hand Furnstahl {\it et al.} \cite{fst} eliminate the scalar chiral partner of
the pion by using the non-linear model and consider that a separate low-mass
scalar generates the intermediate range attraction. They also
incorporate broken scale invariance and obtain an effective Lagrangian
which gives a good account of nuclear matter and finite nuclei.
(The experimental situation in the scalar sector unfortunately remains confused
\cite{scalar}.)

The purpose of the present paper is to further test our Lagrangian by
explicitly
studying pions, since they made no contribution in I. This requires some
extension of the Lagrangian. Firstly, explicit chiral symmetry breaking is
needed to endow the pion with a mass. Secondly, in I the axial coupling
constant,  $g_A$, was unity, rather than the actual value of 1.26; this
defect can be corrected by introducing an additional chiral invariant term
in the Lagrangian, as pointed out by Lee \cite{lee}. Thirdly, if the
isotriplet vector and axial vector mesons (${\rho}$ and
$a_1$) are included in a linear chiral model, along with the scalar and
pseudoscalar mesons, $\pi-a_1$ mixing ensues \cite{gas} and this we explicitly
take into account. The theoretical development is given in Sec. 2, while in
Sec. 3 we give results for nuclei, $\pi N$ scattering and the predicted density
dependence of quantities such as $g_A$ and the pion mass. Our conclusions
are presented in Sec. 4.

\section{Theory}

We begin by recalling the effective Lagrangian employed in I.
It was written in the form
\begin{equation}
{\cal L}_1={\cal L}_0-V_G\;,
\end{equation}
where we have schematically separated out the scale and chiral invariant
part, ${\cal L}_0$, from the potential $V_G$ which induces the breaking of
scale and chiral invariance. We write ${\cal L}_0$ in terms of the nucleon
field $N$, the chiral-invariant combination of sigma and
pi fields, $\sigma$ and $\vmg{\pi}$, the glueball field $\phi$ and
the field of the omega vector meson $\omega_{\mu}$ which is a chiral
singlet. Specifically
\begin{eqnarray}
&&\fpj{\cal L}_0=\thalf\partial_{\mu}\sigma\partial^{\mu}
\sigma+\thalf\partial_{\mu}\vmg{\pi}\cdot\partial^{\mu}\vmg{\pi}
+\thalf\partial_{\mu}\phi\partial^{\mu}\phi-\textstyle{\frac{1}{4}}
\omega_{\mu\nu}\omega^{\mu\nu}
+\thalf \omega_{\mu}\omega^{\mu}G_{\omega\phi}\phi^2\nonumber\\
&&\quad+[(G_4)^2\omega_{\mu}\omega^{\mu}]^2
+\bar N\bigl[\gamma^{\mu}(i\partial_{\mu}-g_{\omega}\omega_{\mu})
-g(\sigma+i\vmg{\pi}\cdot\vmg{\tau}\gamma_5)\bigr]N\;. \label{inv}
\end{eqnarray}
Here the field strength tensor is defined in the usual way,
$\omega_{\mu\nu}=\partial_{\mu}\omega_{\nu}-\partial_{\nu}\omega_{\mu}$
and we have included a term quartic in the vector meson field (note that
since the coefficient of this term is positive, the violation of causality
discussed in Ref. \cite{velo} does not arise here).
In principle the scale invariant $\omega$ mass term could be generated by
coupling to the $\phi$ or the $\sigma$ field, however the latter choice
does not yield a good phenomenology for nuclei \cite{glu3,fs2,fs3} so we
fix on the form $\omega_{\mu}\omega^{\mu}\phi^2$. The vacuum $\omega$ mass
is then $m_{\omega}=G^{\frac{1}{2}}_{\omega\phi}\phi_0$, where
$\phi_0$ denotes the vacuum glueball field (similarly $\sigma_0$ will be
used for the sigma field).

For the potential $V_G$  we used \cite{glu2}
\begin{eqnarray}
V_G(\phi,\sigma,\vmg{\pi})=B\phi^4\left(\ln\frac{\phi}{\phi_0}-\tquar\right)
&\pj-&\pj\thalf B\delta
\phi^4\ln\frac{\sigma^2+\vmg{\pi}^2}{\sigma_0^2}\nonumber\\
&&\quad+\thalf B\delta \zeta^2\phi^2\!\!\left(\sigma^2+\vmg{\pi}^2
-\frac{\phi^2}{2\zeta^2}\right)\;, \label{vg}
\end{eqnarray}
where $\zeta=\frac{\phi_0}{\sigma_0}$ and $B$ and $\delta$ are parameters.
Here the logarithmic terms
contribute to the trace anomaly: in addition to the standard contribution
from the glueball field \cite{mig,gomm} there is also a contribution from
the $\sigma$ field. Specifically the trace
of the ``improved" energy-momentum tensor is
\begin{equation}
\theta^{\mu}_{\mu}=4V_G(\Phi_i)-\sum_i\Phi_i\frac{\partial V_G}{\partial\Phi_i}
=4\epsilon_{\rm vac}\left(\frac{\phi}{\phi_0}\right)^{\!4}\;,
\end{equation}
where $\Phi_i$ runs over the scalar fields \{$\phi,\sigma,\vmg{\pi}$\} and
the vacuum energy, $\epsilon_{\rm vac}=-\tquar B\phi^4_0(1-\delta)$.
Guidance on the value of $\delta$ is provided by the QCD trace anomaly
which is proportional to the beta function. At the one loop level,
with $N_c$ colors and $n_f$ flavors, this is given  by
\begin{equation}
\beta(g)=-\frac{11N_cg^3}{48\pi^2}\left(1-\frac{2n_f}{11N_c}\right)
+{\cal O}(g^5)\;,
\end{equation}
where the first number in parentheses arises from the (antiscreening)
self-interaction of the gluons and the second, proportional to $n_f$, is the
(screening) contribution of quark pairs. This suggests a value of
$\delta=4/33$ for the present case with $n_f=2$ and $N_c=3$ and
we use this here, as in I, although nuclear properties are not
sensitive to the precise value. The third term in the expression for $V_G$
in Eq. (\ref{vg}) is needed to ensure that in the vacuum
$\phi=\phi_0$, $\sigma=\sigma_0$ and $\vmg{\pi}=0$, {\it i.e.} it provides
for spontaneously broken chiral symmetry.

\subsection{Explicit Chiral Symmetry Breaking}

In order to generate a finite pion mass it is necessary to include
explicit chiral symmetry breaking terms in the Lagrangian. To restrict the
possibilities we recall $SU(2)$ symmetry breaking at the quark level
which is generated by ${\cal V}_{SB}=\hat{m}(\bar{u}u+\bar{d}d)$, where
$\hat{m}$ is an average mass. This gives a contribution
$\langle {\cal V}_{SB}\rangle$
to the trace of the energy-momentum tensor, since the operator is
of dimension 3. We require our hadronic potential to have similar dimension.
The double commutator of ${\cal V}_{SB}$ with the axial charge
$[Q^{5a},[Q^{5b},{\cal V}_{SB}]]={\cal V}_{SB}\delta_{ab}$.
We require an analogous
relation to be satisfied in operator form for our hadronic potential.

{}From Eq. (\ref{inv}) the axial current and charge are
\begin{equation}
A^a_{\mu}=-\bar{N}\gamma_{\mu}\gamma_5\thalf\tau^aN
+\pi^a\partial_{\mu}\sigma-\sigma\partial_{\mu}\pi^a\ ;\
Q^{5a}=\int A^a_0(x)d^3x\;.
\end{equation}
It is straightforward to verify that there are three forms for the
symmetry breaking which satisfy our criteria and we write the
general potential
\begin{equation}
V_{SB}=-\epsilon_1\sigma\phi^2-\epsilon_2\sigma(\sigma^2+\vmg{\pi}^2)+
\epsilon_3\bar{N}N\;.
\end{equation}
In order to ensure that in the vacuum $\phi=\phi_0$, $\sigma=\sigma_0$
and $\vmg{\pi}=0$ we need to add additional non-symmetry breaking terms
to $V_{SB}$; we also subtract the vacuum value of these pieces.
It is convenient to set
$\epsilon_1'=\epsilon_1\sigma_0\phi_0^2$,
$\epsilon_2'=\epsilon_2\sigma_0^3$ and work with the ratio of the fields to
their vacuum values, $\chi=\phi/\phi_0$ and
$\nu=\sigma/\sigma_0$. Then
\begin{eqnarray}
&&\fpj V_{SB}'=-\tquar\epsilon_1'\chi^2\left[4\nu-2\left(\frac{\sigma^2
+\vmg{\pi}^2}{\sigma_0^2}\right)-\chi^2\right]\nonumber\\
&&\quad-\tquar\epsilon_2'\left[(4\nu-6\chi^2)\left(
\frac{\sigma^2+\vmg{\pi}^2}{\sigma_0^2}\right)+3\chi^4\right]
+\epsilon_3\bar{N}N-\ttquar(\epsilon_1'+\epsilon_2')\;.\label{vsb}
\end{eqnarray}
Notice that $V_{SB}'$ gives a (relatively small) contribution to the trace of
the energy-momentum tensor. Specifically the contribution to the vacuum
energy is
\begin{equation}
\epsilon_{{\rm vac}}'=-(\epsilon_1'+\epsilon_2')=
-(f_{\pi}m_{\pi})^2\;,
\end{equation}
where for the latter equality, involving the pion decay constant and mass,
we have used the relations given in Subsec. 2.4 below. This is the standard
result for the vacuum energy arising from the symmetry breaking term.

Using the Lagrangian ${\cal L}_1-V_{SB}'$, it is appropriate to give
Lagrange's equations for nuclei or nuclear matter at this point,
since $\vmg{\pi}=0$ in the mean field approximation.
For the nucleon, the Dirac equation is of standard form \cite{horse}, but
now the effective mass is $M^*=g\sigma+\epsilon_3=M\nu+\epsilon_3(1-\nu)$.
The equations for the fields $\chi(r)=\phi(r)/\phi_0$,
$\nu(r)=\sigma(r)/\sigma_0$ and $\omega_0(r)$ can be written
\begin{eqnarray}
&&\fpj\phi_0^2{\cal D}\chi-2[B_0(2-\delta)+\epsilon_1'-3\epsilon_2']\chi
+2[B_0\delta-3\epsilon_2']\nu\nonumber\\
&&=4B_0[\chi^3(\ln\chi-\delta\ln\nu)-\chi+\delta\nu]
+B_0\delta[\chi(\nu^2-\chi^2)+2(\chi-\nu)]\nonumber\\
&&\ -m_{\omega}^2\omega_0^2\chi-\epsilon_1'(2\chi\nu-\chi^3-\chi\nu^2+2\chi)
-3\epsilon_2'[\chi^3-\chi\nu^2-2\chi+2\nu]\;,\nonumber\\
&&\fpj\sigma_0^2{\cal D}\nu-(2B_0\delta+\epsilon_1'-3\epsilon_2')\nu
+2(B_0\delta-3\epsilon_2')\chi\nonumber\\
&&=g\sigma_0\rho_s-B_0\delta\left[\frac{\chi^4}{\nu}-\chi^2\nu
+2(\nu-\chi)\right]-\epsilon_1'(\chi^2-\chi^2\nu+\nu)\nonumber\\
&&\qquad-3\epsilon_2'(\nu^2-\chi^2\nu+2\chi-\nu)\;,\nonumber\\
&&\fpj {\cal D}\omega_0-m_{\omega}^2\omega_0=-g_{\omega}\rho_B+m_{\omega}^2
(\chi^2-1)\omega_0+4(G_4)^4\omega_0^3\;,\label{fieldeq}
\end{eqnarray}
where the densities $\rho_s=\langle\bar{N}N\rangle$ and
$\rho_B=\langle N^{\dagger}N\rangle$ can be expressed in terms
of the components of the nucleon Dirac spinors in the usual way \cite{horse}.
In eq. (\ref{fieldeq}) we have made the definitions
${\cal D}\equiv\frac{d^2}{dr^2}
+\frac{2}{r}\frac{d}{dr}$ and $B_0\equiv B\phi_0^4$. The terms linear in the
fields, {\it i.e.}, the kinetic energy and mass terms, have been separated
out on the left of these equations. In the case of
nuclear matter the fields are constant and so the derivatives are
zero. We note that the mass matrix is not diagonal and in
order to solve the equations for nuclei it is necessary to go to a
representation which is diagonal in the limit that $r\rightarrow\infty$,
{\it i.e.} $\chi,\nu\rightarrow1$. This is discussed in the Appendix A.

The energy-momentum tensor can be used to obtain the total energy of the
system in the standard way \cite{sew}. Subtracting constants so that the
energy is measured relative to the vacuum, we obtain
\begin{eqnarray}
E&\pj=&\pj\sum_{\alpha}^{\rm occ}\epsilon_{\alpha}(2j_{\alpha}+1)
-2\pi\int_0^{\infty}dr\:r^2\bigg\{g\sigma_0\nu\rho_s+g_{\omega}\omega_0\rho_B
\nonumber\\
&&+2B_0[\chi^4(\ln\chi-\delta\ln\nu+\tquar)-\tquar]+\thalf B_0\delta
[\chi^2(2\nu^2-3\chi^2)+1]\nonumber\\
&&\quad-m_{\omega}^2\chi^2\omega_0^2-2(G_4)^4\omega_0^4+\thalf\epsilon_1'
[\chi^2(\chi^2-2\nu+2\nu^2)-1]\nonumber\\
&&\quad\qquad+\thalf\epsilon_2'(6\nu^2\chi^2-3\chi^4-2\nu^3-1)\bigg\}\;.
\label{energy}
\end{eqnarray}
In the first term on the right the $\epsilon_{\alpha}$ are the Dirac single
particle energies and $j_{\alpha}$ is the total angular momentum of the single
particle state. In nuclear matter this term becomes
$4\sum_{\vm{k}}\left(g_{\omega}\omega_0+\sqrt{k^2+M^{*2}}\right)$ and the
integral in Eq. (\ref{energy}) is trivial since the fields are constant.

Finally, for
nuclei the photon field is included in the standard way \cite{horse}
and, since the contribution from the $\rho$ field is small, it is handled
in the simplest manner as in I. These fields are suppressed
in the above equations.

\subsection{Axial Coupling Constant}

Eqs. (\ref{inv}) and (\ref{vsb}) yield, in the vacuum,
$g\sigma_0=M-\epsilon_3$ as the form of the
Goldberger-Treiman relation \cite{gold,cam} which corresponds to a nucleon
axial coupling constant $g_A=1$. It has been pointed out by Lee \cite{lee}
that this defect can be corrected by introducing an additional chiral invariant
term in the Lagrangian. This has also been discussed by Akhmedov \cite{ak} in
the context of the standard chiral model, which, as we have remarked, is
unable to reproduce the observed properties of finite nuclei.

It is straightforward to motivate the additional term. With the definitions
\begin{equation}
\Sigma=\sigma+i\vmg{\pi}\cdot\vmg{\tau}\quad;\quad
\Pi=\sigma+i\vmg{\pi}\cdot\vmg{\tau}\gamma_5\;,\label{SigPi}
\end{equation}
one observes that
\begin{equation}
\bar{N}\gamma_{\mu}\Pi^{\dagger}\left(\partial^{\mu}\Pi\right) N=
\bar{N}_R\gamma_{\mu}\Sigma^{\dagger}\left(\partial^{\mu}\Sigma\right)N_R
+\bar{N}_L\gamma_{\mu}\Sigma\left(\partial^{\mu}\Sigma^{\dagger}\right)N_L\;,
\label{newch}
\end{equation}
where $N_{L,R}$ are the nucleon solutions of left- and right-handed chirality.
Under global $SU_L(2)\times SU_R(2)$ transformations
$\Sigma\rightarrow L\Sigma R^{\dagger}$, $N_L\rightarrow LN_L$ and
$N_R\rightarrow RN_R$, so the expression (\ref{newch}) is clearly invariant.

Simply Hermitizing the expression (\ref{newch}) yields
$\bar{N}\gamma_{\mu}\partial^{\mu}\left(\Pi^{\dagger}\Pi\right)N$ and, since
the nucleon vector current is conserved, this can be written as a total
divergence which gives no contribution to the action. A more interesting
possibility \cite{lee} is
\begin{eqnarray}
{\cal L}_2&\pj=&\pj \frac{D}{4\phi^2}\bar{N}\gamma_{\mu}i\left[
\left(\partial^{\mu}\Pi^{\dagger}\right)\Pi-\Pi^{\dagger}
\left(\partial^{\mu}\Pi\right)\right]N\nonumber\\
&\pj=&\pj \frac{D}{\phi^2}\bar{N}\gamma_{\mu}\thalf\vmg{\tau}\cdot\left[
\vmg{\pi}\times\partial^{\mu}\vmg{\pi}
+\gamma_5\left(\sigma\partial^{\mu}\vmg{\pi}-\vmg{\pi}\partial^{\mu}\sigma
\right)\right]N\;,\label{newl2}
\end{eqnarray}
where we have inserted a constant $D$ and a factor of $\phi^{-2}$ to ensure
scale invariance. Since in a mean field calculation of nuclei or nuclear
matter $\vmg{\pi}=0$, ${\cal L}_2$ will not modify the results obtained in I.
It will, however,  contribute to the axial vector
coupling constant, $g_A$, and the $\pi NN$ coupling, $g_{\pi NN}$, so that
\cite{lee,ak} the Goldberger-Treiman relation
$g_{\pi NN}\sigma_0=g_A M-\epsilon_3$ is obtained. Before giving
further details, the renormalization of the pion field needs to be
addressed.

\subsection{$\pi-a_1$ Mixing}

The inclusion of the isotriplet vector and axial vector mesons (${\rho}$ and
$a_1$), along with the scalar and pseudoscalar mesons, in a linear chiral model
results in significant complication. The details are
discussed by Ko and Rudaz \cite{ko}  (see also Gasiorowicz and Geffen
\cite{gas}) and their vector meson Lagrangian is given in Appendix B.
When the vector mesons are included, the principle of vector meson
dominance in the currents is implemented by requiring that, apart from
explicit symmetry breaking terms, the
Lagrangian be invariant under local $SU_L(2)\times SU_R(2)$ save for the
${\rho}$ and $a_1$ mass terms . This requires the replacement of the
ordinary derivatives of the $\sigma$ and $\vmg{\pi}$ fields in Eq.
(\ref{inv}) by covariant derivatives:
\begin{eqnarray}
\partial_{\mu}\sigma&\pj\rightarrow&\pj\Delta_{\mu}\sigma=\partial_{\mu}\sigma
+f\vm{a}_{\mu}\cdot\vmg{\pi}\;,\nonumber\\
\partial_{\mu}\vmg{\pi}&\pj\rightarrow&\pj\Delta_{\mu}\vmg{\pi}
=\partial_{\mu}\vmg{\pi}
+f\vmg{\rho}_{\mu}\times\vmg{\pi}-f\sigma\vm{a}_{\mu}\;,\label{deriv}
\end{eqnarray}
where $f$ is the gauge coupling constant for the vector fields.
This induces mixing  between the $\pi$ and the
$a_1$ which results in a renormalization of the pion field.
Considering just the kinetic energy
terms for the $\pi$ and $a_1$, the mass term for the $a_1$ and the mixing,
we have
\begin{eqnarray}
&&\fpj\thalf\partial_{\mu}\vmg{\pi}\cdot\partial^{\mu}\vmg{\pi}
-\tquar\vm{a}_{\mu\nu}\cdot\vm{a}^{\mu\nu}
+\thalf m_a^{*2}\vm{a}_{\mu}\cdot\vm{a}^{\mu}
-f\sigma\partial_{\mu}\vmg{\pi}\cdot\vm{a}^{\mu}\nonumber\\
&&\fpj\rightarrow\thalf Z^*_{\pi}\partial_{\mu}\vmg{\pi}\cdot\partial^{\mu}
\vmg{\pi} -\tquar\vm{a}'_{\mu\nu}\cdot\vm{a}'^{\mu\nu}
+\thalf m_a^{*2}\vm{a}'_{\mu}\cdot\vm{a}^{\prime\mu}\;,\label{mix}
\end{eqnarray}
where $\vm{a}_{\mu\nu}=\partial_{\mu}\vm{a}_{\nu}-\partial_{\nu}\vm{a}_{\mu}$,
the transformed field is
\begin{equation}
\vm{a}'_{\mu}=\vm{a}_{\mu}-\frac{f\sigma}{m_a^{*2}}
\partial_{\mu}\vmg{\pi}\;,\label{moda}
\end{equation}
and
\begin{equation}
Z^*_{\pi}=1-(f\sigma/m^*_a)^2\;.\label{zpia}
\end{equation}
Here, and in the following, an asterisk indicates that the values of the
$\sigma$ and $\phi$ fields in nuclear matter are used; thus
$m^*_a$ denotes the mass of the $a_1$ in matter.
Guided by the observation in I that nuclei prefer the $\rho$ mass to
be generated by coupling to the glueball field rather than the $\sigma$
field, we take the constants $b$ and $c$ of Ref. \cite{ko} to be equal so
that the masses in matter are
\begin{equation}
m^{*2}_a=m^{*2}_{\rho}+(2c+1)(f\sigma)^2\quad;\quad
m^{*2}_{\rho}=m^{2}_{\rho}\left(\frac{\phi}{\phi_0}\right)^{\!2}\;,
\label{massrhoa}
\end{equation}
where the absence of an asterisk on the mass implies the vacuum value.
Here we have implicitly assumed that the fields are
replaced by their average values in nuclear matter, {\it i.e.} used the mean
field approximation, in which case $Z^*_{\pi}$ is a renormalization
and the physical pion field is defined by
$\vmg{\pi}^r=\sqrt{Z^*_{\pi}}\vmg{\pi}$.

The redefinition (\ref{moda}) of the $a_1$ field in the Lagrangian
(\ref{lvector}) of Appendix B and in the $a_1N$ coupling,
$-\thalf f\bar{N}\gamma^{\mu}\gamma_5\vmg{\tau}N\cdot\vm{a}_{\mu}$,
leads to additional terms involving the $\sigma$ and $\vmg{\pi}$
fields. We can neglect those which are ${\cal O}(\vmg{\pi}^4)$ or
${\cal O}(\partial\sigma\partial\vmg{\pi})^2$ since they will not
contribute here. The additional terms we need to consider are
\begin{equation}
{\cal L}_3=\thalf(Z^*_{\pi}-1)\left[\partial_{\mu}\vmg{\pi}\cdot
\partial^{\mu}\vmg{\pi}+\sigma^{-1}\bar{N}\gamma^{\mu}\gamma_5\vmg{\tau}\cdot
N\partial_{\mu}\vmg{\pi}\right]
+\frac{f^2\sigma}{m^{*2}_a}\partial^{\mu}\sigma
\vmg{\pi}\cdot\partial_{\mu}\vmg{\pi}\,,\label{newl3}
\end{equation}
where we have included the renormalization of Eq. (\ref{mix}).
Again, these terms will not affect nuclear matter at the mean field level.

We must also replace the derivatives in Eq. (\ref{newl2}) by covariant
derivatives yielding the modified form
\begin{eqnarray}
{\cal L}_2'&\pj=&\pj\frac{D}{\phi^2}\bar{N}\gamma_{\mu}\thalf\vmg{\tau}
\cdot\Bigl[Z_{\pi}^*\vmg{\pi}\times\partial^{\mu}\vmg{\pi}
+\gamma_5\Bigl(Z^*_{\pi}\sigma\partial^{\mu}\vmg{\pi}
-\vmg{\pi}\partial^{\mu}\sigma\nonumber\\
&&\qquad\qquad\qquad\qquad-(1-Z^*_{\pi})\sigma^{-1}\vmg{\pi}
(\vmg{\pi}\cdot\partial^{\mu}\vmg{\pi})\Bigr)\Bigr]N\;.\label{newl2p}
\end{eqnarray}
Thus our total Lagrangian is
\begin{equation}
{\cal L}_{\rm total}={\cal L}_1+{\cal L}_2'+{\cal L}_3-V_{SB}'\;.
\label{ltotal}
\end{equation}

\subsection{Effective Couplings and the Pion Mass}

The axial current, $\vm{A}_{\mu}$, arises from the
breaking of local $SU_L(2)\times SU_R(2)$ by the $\rho$ and $ a_1$ mass
terms and can be written \cite{ko}
\begin{eqnarray}
\vm{A}_{\mu}&\pj=&\pj-f^{-1}\left[m^{*2}_a+f^2(c\vmg{\pi}^2-\sigma^2)
\right]\vm{a}_{\mu}\nonumber\\
&&\qquad\qquad-cf\left[2\sigma\vmg{\pi}\times\vmg{\rho}_{\mu}
+2(\vm{a}_{\mu}\cdot\vmg{\pi})\vmg{\pi}-\vmg{\pi}^2\vm{a}_{\mu}\right]\;.
\label{axcur}
\end{eqnarray}
Using this, we can determine the axial vector coupling constant in medium,
$g_A^*$, by a tree-level calculation of
$\langle N(k-q)|\vm{A}_{\mu}|N(k)\rangle$ and identification
of the coefficient of $-\bar{N}\gamma_{\mu}\gamma_5\thalf\vmg{\tau}N$ in
the limit $q\rightarrow0$. This gives
\begin{equation}
g_A^*=Z^*_{\pi}\left(1+\frac{D\sigma^2}{\phi^2}\right)\;.\label{ga}
\end{equation}
Since in the vacuum $Z_{\pi}<1$, the presence of the new Lagrangian
${\cal L}_2'$ is essential if one is to obtain $g_A>1$.

Turning to the effective $\pi NN$ coupling
$-ig^*_{\pi NN}\bar{N}\gamma_5\vmg{\tau}\cdot\vmg{\pi}^rN$, where
$\vmg{\pi}^r$ denotes the renormalized (physical) pion field, we find
\begin{equation}
g^*_{\pi NN}=\left(\sigma\sqrt{Z^*_{\pi}}\right)^{-1}\left[g\sigma+M^*(g^*_A-1)
\right]\;.\label{gpinn}
\end{equation}
Here we have used the relation
\begin{equation}
\bar{N}\gamma_{\mu}\gamma_5\vmg{\tau}\cdot\partial^{\mu}\vmg{\pi}N=
-2iM^*\bar{N}\gamma_5\vmg{\tau}\cdot\vmg{\pi}N\;,
\end{equation}
for on mass-shell nucleons.

Using Eqs. (\ref{moda}) and (\ref{axcur}), we have
\begin{equation}
\langle0|A^b_{\mu}(0)|\pi^{ra}(k)\rangle=i\sigma\sqrt{Z^*_{\pi}}
\,k_{\mu}\delta_{ab}\;,
\end{equation}
so that the effective pion decay constant in medium is
$f_{\pi}^*=\sigma\sqrt{Z^*_{\pi}}$ (we use the standard nomenclature here
while noting that there would be additional contributions to the
calculation of the decay of an actual pion in medium).
Then Eqs. (\ref{ga}) and (\ref{gpinn})
yield the Goldberger-Treiman \cite{gold} relation in medium
\begin{equation}
g^*_{\pi NN}f^*_{\pi}=g_A^*M^*-\epsilon_3\;,
\end{equation}
which is a consequence of the chiral symmetry employed.

Finally we discuss the pion mass in symmetric (isospin zero) nuclear matter.
In addition to the contributions that
are generated by straightforward application of Lagrange's equations to
${\cal L}_{\rm total}$ of Eq. (\ref{ltotal}), one  must include \cite{mat}
the fluctuations in the axial density of nucleons that are induced by the
presence of pions. This is represented by the propagator contribution in
Fig. 1. We choose to evaluate the contribution to the pion mass at
four-momentum $q=0$ (rather than identifying the pole of the propagator
at $\vm{q}=0$) which yields $g^2\rho_s/(Z_{\pi}^*M^*)$. In this
approximation the effective pion mass is given by
\begin{equation}
m_{\pi}^{*2}=\frac{1}{Z_{\pi}^*\sigma_0^2}\left[\frac{\epsilon_1'\chi^2}
{\nu}+\epsilon_2'\nu-\frac{\epsilon_3\rho_s}
{\nu(\nu+\epsilon_3/(g\sigma_0))}\right]\;.\label{mpi}
\end{equation}
In the vacuum this evidently becomes
\begin{equation}
m_{\pi}^{2}=\frac{\epsilon_1'+\epsilon_2'}{Z_{\pi}\sigma_0^2}\;.
\label{mpivac}
\end{equation}
We can now evaluate the pion-nucleon sigma term, which with our symmetry
breaking is simply $\langle N(p')|V_{SB}|N(p)\rangle$, evaluated at
$t=(p-p')^2=0$. In the tree approximation, following the approach of
Campbell \cite{cam}, we obtain
\begin{equation}
\Sigma(0)=\frac{2g\sigma_0Z_{\pi}}{m_<^2m_>^2}\left[3m_{\pi}^2\gamma
+\frac{2\epsilon_1'}{Z_{\pi}\sigma_0^2\phi_0}(\beta\sigma_0-\gamma\phi_0)
\right]+\epsilon_3\;,\label{sigmat}
\end{equation}
using the definitions of Appendix A.

\section{Results}

\subsection{Nuclei}

We shall examine the results obtained with and without the
$(\omega_{\mu}\omega^{\mu})^2$ term in the Lagrangian. For the latter case,
the binding energy/particle (16 MeV) and density
($\rho_B^{\rm sat}=0.148$ fm$^{-3}$) of
equilibrium nuclear matter are sufficient to fix the parameters $g_{\omega}$
and $B_0$ for various assumed values of the symmetry breaking constants
$\epsilon_1'$, $\epsilon_2'$ and $\epsilon_3$. For the former,
an additional parameter enters and we take $G_4/g_{\omega}=0.19$. This
is similar to the value used in I (where the glueball field was frozen,
$\chi=1$) and is judged to give the best results for
the nuclear properties; smaller values of $G_4$ produce little effect, whereas
significantly larger values preclude a fit to equilibrium nuclear matter.
As in I, two solutions are found and we pick that which yields the most
reasonable effective mass, $M^*_{\rm sat}$, and compression modulus, $K$.
The parameters are listed in Table 1 for the cases where $G_4=0$ and $G_4>0$.
These parameters actually correspond to our favored values for the symmetry
breaking, with $\epsilon_1'$ chosen to give the pion mass
(Eq. (\ref{mpivac})) for $\epsilon_2'=0$, and $\epsilon_3=-15$ MeV.
However the quantities listed are insensitive to the values of these
parameters since the scale of the symmetry breaking is small in comparison to
the other scales involved.

The first few quantities in Table 1 are similar to those discussed in I.
Thus the vacuum energy is in rather good agreement with QCD sum rule
estimates \cite{svz} of $(240\ {\rm MeV})^4$, as in I, while the $\omega$
coupling is comparable to that needed for the nucleon-nucleon interaction
\cite{bonn}. Since nuclei are not sensitive to the ratio
$\zeta=\phi_0/\sigma_0$ we fix the value such that the larger scalar mass
$m_>=1.5$ GeV in view of QCD sum rule estimates \cite{shif} for a dominantly
glueball state. The lower mass $m_<\simeq0.5$ GeV. The saturation effective
mass ratio agrees with the recent QCD sum rule estimate
\cite{lein} of $0.73\pm0.09$. As discussed in I, the lower effective mass
for $G_4>0$ improves the spin-orbit splittings in nuclei which are about
80\% of the observed values, versus 65\% when $G_4=0$.
Reference \cite{lein} also determines the vector
self-energy of a nucleon in nuclear matter, $g_{\omega}\omega_0$, to be
$0.31\pm0.06$ GeV. We obtain 0.22 GeV with $G_4=0$ and 0.28 GeV for $G_4>0$
which suggests that the latter yields the better phenomenology.
As regards the compression modulus, $K$, current estimates are in the range
200--300 MeV \cite{comp} which again favors the case with $G_4>0$. Defining
the third derivative of the binding energy per particle as
$S=k^3_F\frac{d^3}{dk_F^3}\left(\frac{E}{A}\right)$ (evaluated at
equilibrium), we give in  Table 1 the ratio $S/K$ since
the data indicate \cite{pear} a linear relation between this quantity and
$K$. For $G_4>0$ our point lies within the allowed error band, while for
$G_4=0$ it is slightly off the band.

For nuclei we need to fix $\sigma_0$ and the values which give the best
fit are listed in Table 1; a reduction in these values
leads to poor agreement, but a modest increase does not
substantially change the results presented in this section. The vacuum pion
renormalization, $Z_{\pi}$, and the gauge coupling, $f$ can then
be determined. Since the latter is the $\rho NN$
coupling constant, it can be compared with the value of 5.3$\pm$0.3 given
by Dumbrajs {\it et al.} \cite{dum}; our results are of
reasonable magnitude. The remaining quantities in Table 1 will be referred
to below.

Since the single particle levels and charge density distributions are quite
similar to those obtained in I, we shall content ourselves with discussing
the bulk properties of oxygen, calcium and lead. These are given in Table 2;
here the appropriate corrections for c.m.
\cite{neg} and finite size effects \cite{horse} have been made.
We also list the value of the sigma term from Eq. (\ref{sigmat}).
With the choice $g_{\pi NN}^2/(4\pi)=14.3$ from H\"ohler \cite{ho},
the vacuum axial coupling contant, $g_A$, is determined and the values
in Table 2 are seen to be quite reasonable.
The parameter sets designated with a W in Table 2 correspond to the case
with $G_4>0$. The cases labelled A correspond to no explicit
symmetry breaking so, while $g_A$ is reasonable, the sigma term is
necessarily zero; the remaining results are similar to those given in I.
The other cases show the effect of various
symmetry breaking parameters; here $\epsilon_1'$ and $\epsilon_2'$
are constrained by Eq. (\ref{mpivac}) to give the vacuum pion mass.

The symmetry breaking tends to decrease the binding energy and increase the
radius or {\it vice versa}, but, as expected, the effects are fairly small.
Simply choosing $\epsilon_1'\neq0$ (case WB) improves the radii at the
expense of some loss of binding. Also the sigma term becomes non-zero and
is reasonable in comparison to the estimate of
Gasser {\it et al.} \cite{gass} of 45 MeV with $\sim$15\% error.
If we include $\epsilon_2'$ (case WC), the radii are reduced and the sigma
term becomes rather large, although the latter could be corrected
by an appropriate choice of $\epsilon_3$.
We have also examined the choice $\epsilon_2'=-\thalf\epsilon_1'$,
but this significantly increase $K$ and correspondingly substantially
reduces the binding energies. We also note that with $\epsilon_2'\neq0$,
the $\sigma^3$ term in $V_{SB}$ would dominate over the potential $V_G$ as
$\sigma\rightarrow\infty$, which would seem to be unphysical. We therefore
favor setting $\epsilon_2'=0$. With this choice, we show in Table 2
the results obtained with various values of $\epsilon_3$.
A small negative value of $\simeq-15$ MeV appears to give the best
results for nuclei and also for the sigma term. Set WF gives rather a good
account of the radii with binding energies that are somewhat low, while set
F improves the binding energies at the expense of the radii.

\subsection{$\pi$N Scattering}

The Lorentz-invariant scattering matrix is conventionally written in the
form $\bar{N}(p_2)T_{ba}N(p_1)$, where
\begin{equation}
T_{ba}=\left[A^{(+)}+\thalf(\qsl_1+\qsl_2)B^{(+)}\right]\delta_{ba}
+\left[A^{(-)}+\thalf(\qsl_1+\qsl_2)B^{(-)}\right]i\epsilon_{bac}
\tau^c\:.
\end{equation}
Here $q_1,\ a$ ($q_2,\ b$) specify the four momentum and isospin component
of the incoming (outgoing) meson. It is straightforward, though tedious,
to obtain the amplitudes in terms of the usual Mandelstam variables
$s=(p_1+q_1)^2$, $t=(q_1-q_2)^2$ and $u=(p_1-q_2)^2$. The plus
amplitudes receive contributions from nucleon exchange in the $s$ and $u$
channels and from the exchange of scalar mesons in the $t$ channel. With
on-mass-shell pions, we find
\begin{eqnarray}
A^{(+)}&\pj=&\pj-\frac{2g}{\sigma_0Z_{\pi}(t-m_>^2)(t-m_<^2)}\Biggl\{
\thalf Z_{\pi}(t-2\gamma)(t+m_{\pi}^2)\nonumber\\
&&+\left(t-2\gamma+2\beta\zeta^{-1}\right)\left[(2m_{\pi}^2-\thalf t)
\frac{f^2m_{\rho}^2\sigma_0^2}{m_a^4}-\frac{\epsilon_1'}{\sigma_0^2}
\right]\Biggr\}\nonumber\\
&&+\frac{g_{\pi NN}^2}{M}+\frac{g\epsilon_3}{Z_{\pi}\sigma_0M}\;,\nonumber\\
B^{(+)}&\pj=&\pj g_{\pi NN}^2\left(\frac{1}{u-M^2}-\frac{1}{s-M^2}\right)\;.
\label{abplus}
\end{eqnarray}
Here $m_>$ and $m_<$ denote the scalar meson masses which are given in the
Appendix A, along with the definitions for $\alpha,\beta$ and $\gamma$.
Notice that $B^{(+)}$ is of standard form \cite{sew,mat}.

For the minus amplitudes, nucleon exchanges in the $s$ and $u$ channels
again contribute and in the $t$ channel we need to consider $\rho$ exchange.
For the latter, the $\rho\pi\pi$ coupling follows from the Lagrangian
quoted in Appendix B and is given by Ko and Rudaz \cite{ko} as
\begin{eqnarray}
{\cal L}_{\rho\pi\pi}&\pj=&\pj-\frac{fm_{\rho}^2}{Z_{\pi}m_a^2}
\vmg{\rho}_{\mu}\cdot(\partial^{\mu}\vmg{\pi}^r\times\vmg{\pi}^r)\nonumber\\
&&\qquad-\left(\frac{f^3\sigma_0^2}
{2Z_{\pi}m_a^4}+\frac{fZ_{\pi}\kappa_6}{2m_{\rho}^2}\right)\vmg{\rho}_{\mu\nu}
\cdot(\partial^{\mu}\vmg{\pi}^r\times\partial^{\nu}\vmg{\pi}^r)\;,
\label{rhopipi}
\end{eqnarray}
where $\vmg{\rho}_{\mu\nu}=\partial_{\mu}\vmg{\rho}_{\nu}-
\partial_{\nu}\vmg{\rho}_{\mu}$.

The $\rho NN$vertex is
\begin{equation}
{\cal L}_{\rho NN}=\thalf f\bar{N}\gamma^{\mu}\vmg{\tau}\cdot
\vmg{\rho}_{\mu}N\;. \label{rhonn}
\end{equation}
In addition we have a four point $NN\pi\pi$ interaction in Eq. (\ref{newl2p})
which must be considered. The resulting amplitudes are
\begin{eqnarray}
A^{(-)}&\pj=&\pj0\;,\nonumber\\
B^{(-)}&\pj=&\pj-g_{\pi NN}^2\left(\frac{1}{u-M^2}+\frac{1}{s-M^2}\right)
+\frac{2Z_{\pi}-g_A^2-1}{2Z_{\pi}\sigma_0^2}\nonumber\\
&&-\frac{f^2}{Z_{\pi}m_a^2(t-m_{\rho}^2)}\left[m_{\rho}^2-t\left(
\frac{f^2\sigma_0^2}{2m_a^2}+\frac{\kappa_6Z_{\pi}^2m_a^2}{2m_{\rho}^2}
\right)\right]\;. \label{abminus}
\end{eqnarray}

In terms of these amplitudes, the $s$-wave scattering lengths and effective
ranges, $a_0^{(\pm)}$ and $r_0^{(\pm)}$, and the $p$-wave scatterings
lengths, $a_{1\pm}^{(\pm)}$, are \cite{mat}
\begin{eqnarray}
a_0^{(\pm)}&\pj=&\pj\eta\left(A_0^{(\pm)}+m_{\pi}B_0^{(\pm)}\right)
\quad;\quad a_{1+}^{(\pm)}=\twothird\eta C_0^{(\pm)}\;,\nonumber\\
a_{1-}^{(\pm)}&\pj=&\pj\twothird\eta C_0^{(\pm)}-\frac{\eta}{4M^2}\left[
A_0^{(\pm)}-(2M+m_{\pi})B_0^{(\pm)}\right]\;,\nonumber\\
r_0^{(\pm)}&\pj=&\pj-\eta\Biggl\{2C_0^{(\pm)}-\frac{(M+m_{\pi})^2}{Mm_{\pi}}
D_0^{(\pm)}\nonumber\\
&&+\frac{1}{2Mm_{\pi}}\left[\left(1-\frac{m_{\pi}}{2M}\right)A_0^{(\pm)}
-\left(M+\frac{m_{\pi}^2}{2M}\right)B_0^{(\pm)}\right]\Biggr\}\;,
\label{ars}
\end{eqnarray}
where $C_0=[\partial(A+m_{\pi}B)/\partial t]_{t=0}$,
$D_0=[\partial(A+m_{\pi}B)/\partial s]_{t=0}$, with $s$ and $t$ taken to be
the independant variables. Also $\eta^{-1}=4\pi(M+m_{\pi})/M$.
The subscript 0 implies that the amplitudes are to be evaluated at
threshold:-- $t=0$, $s=(M+m_{\pi})^2$ and $u=(M-m_{\pi})^2$.

In order to calculate the scattering amplitudes in the most general case,
additional parameters are needed. The parameter $D$ is determined
from Eqs. (\ref{ga}) and (\ref{gpinn}) in the vacuum and is listed in Table
1. The quantity $c$ in Table 1 is deduced from the value of $Z_{\pi}$
and Eqs. (\ref{zpia}) and (\ref{massrhoa}) in vacuum.
The parameter $\kappa_6$ listed in Table 1 is obtained by fitting the
decay width, $\Gamma(\rho\rightarrow\pi\pi)=151.2\pm1.2$ MeV \cite{pdg},
using Eq. (\ref{rhopipi}) with the $\rho$ on the mass shell.
The effective $\rho\pi\pi$ coupling constant of 6.05 is of similar
magnitude to the $\rho NN$ coupling $f$ of Table 1, which is in accord
with the notion of $\rho$ universality.
Since significant cancellations may occur in calculating the scattering
lengths and effective ranges, Eq. (\ref{ars}), we prefer to deduce $A_0$,
$B_0$, $C_0$ and $D_0$ from the experimental data \cite{ho} and these
are listed in the first row of Table 3. The  $\Delta$ resonance
contributions to the amplitudes, constructed such that no double counting
occurs, have been given by H\"ohler \cite{ho}. We list these in Table 3,
along with the difference which we would hope to reproduce with our model.
(For definiteness we note that the off-shell parameter $Z$ appearing in the
$\pi N\Delta$ interaction was chosen to be $\thalf$.)
We first give results obtained with our simplest Lagrangian, ${\cal L}_1$.
Since $g_{\pi NN}=g$ in this model, the results are quite poor. If
we also include ${\cal L}_2$, this allows us to fit $g_{\pi NN}$ by
adjusting $D$ (here $Z_{\pi}$ is still 1) and this gives much better
agreement with the data. The remaining cases in Table 3 correspond
to the full model and the notation follows that of Table 2.
With no explicit symmetry breaking (cases A and WA), the sum of $A_0^{(+)}$ and
$m_{\pi}B_0^{(+)}$ must cancel so that there is no contribution to
$a_0^{(+)}$ in the chiral limit.
In fact the sum should be negative, as it is for most of the cases
displayed. The sensitivity to the symmetry breaking is not great, but there
appears to be a weak preference for cases F and WF and we recall that these
were also favored from the results of Table 1.

The question arises as to how well the scattering parameters test our model.
In this connection it is useful to compare with the results of Matsui and
Serot \cite{sew,mat} who used both the  standard chiral model and QHD-II.
In all models $g_{\pi NN}$ is taken from experiment so that the predictions
for $B_0^{(+)}$ and $D_0^{(+)}$ are the same. Also $A_0^{(+)}$ will
only differ from $-m_{\pi}B_0^{(+)}$ because of symmetry breaking terms,
which are necessarily small. In Ref. \cite{sew,mat}, $C_0^{(+)}$ is
2.0 (3.0) in the chiral (QHD-II) model where the scalar mass is 770 (520)
MeV. Thus one distinct feature of the present model is that we get about
the right value for $C_0^{(+)}$ with $m_<\sim 500$ MeV. As regards the $(-)$
amplitudes, $D_0^{(-)}$ is fixed once $g_{\pi NN}$ is given and to a large
extent $C_0^{(-)}$ is also; thus all models give much the same result.
However in QHD-II, where the $\rho$ is included, $m_{\pi}B_0^{(-)}=3.1$,
whereas our value is smaller, due to the constant term and the more
complicated form for the $\rho$ contribution, which brings us closer to the
desired value.

\subsection{Density Dependence of the Parameters}

In our Lagrangian there are two scales, $\chi$ and $\omega$ which give the
ratio of the glueball and sigma fields to their vacuum values. Their
dependence on density is shown in Fig. 2 for the parameters of Table 1.
The glueball field changes very
little which implies that at saturation the $\omega$ and $\rho$ masses are
reduced by only 3\%. By contrast QCD sum rule estimates \cite{qcdvec}
suggest a reduction of roughly $20\pm10$\%. Brown and Rho \cite{scaling}
have also argued for a similar scaling. An effect of this magnitude
would suggest scaling these vector masses with $\nu$ instead of $\chi$.
Then, however, it is not possible \cite {glu3,fs2,fs3} to produce acceptable
fits to the properties of nuclei, for instance the charge density
distributions start to develop oscillations in disagreement with the data
and these probably signal \cite{price} the onset of a situation where the
ground state of nuclear matter is no longer uniform.

The pion decay constant in medium is given by
$f_{\pi}^*/f_{\pi}=\nu\sqrt{Z_{\pi}^*/Z_{\pi}}$. The behavior of the pion
renormalization, $Z_{\pi}^*/Z_{\pi}$, is shown in Fig. 3. The change in
$Z_{\pi}^*$ in going from the vacuum to saturation
is a 7--11\% effect, depending on the parameters chosen, so that
$f_{\pi}^*/f_{\pi}\simeq M^*/M$ as suggested by Brown and Rho
\cite{scaling}. Specifically the pion decay constant ratio at saturation
is 0.74 ($G_4=0$) or 0.66 ($G_4>0$) which compare quite well with the
effective masses given in Table 1. A similar level of agreement is obtained
if one also includes a factor of $\sqrt{g_A^*/g_A}$. The ratio of the axial
coupling constants, given in Fig. 4, shows the expected reduction with
density. The data indicate \cite{quench} that $g_A^*$ in equilibrium nuclear
matter should be close to 1, which is a little lower than our value of 1.1.

The quantities discussed thus far are insensitive to the value of the
explicit symmetry breaking parameters. This is not the case for the
effective pion
mass, however, as we show in Fig. 5. At low densities, where the linear
approximation is sufficient, the behavior is given by the expression
\cite{vest,eric} $m_{\pi}^*/m_{\pi}=1-2\pi a_0^{(+)}\rho_B/(m_{\pi}m_R)$,
where $m_R$ is the reduced mass of the pion-nucleon system. (This can also
be obtained in leading order from Eqs. (\ref{mpi}), (\ref{abplus}) and
(\ref{ars}).) The mass
decreases for set WD since $a_0^{(+)}$ is positive, whereas is should be
negative leading to an increased pion mass in matter. The pionic atom data
indicate a 10\% increase at saturation \cite{eric} and this suggests that
$\epsilon_3$ should be zero or small and negative in agreement with the
previous indications.

\section{Conclusions}

The effective Lagrangian of I, which embodied broken scale and spontaneously
broken chiral symmetry as suggested by QCD, gave a good account of the
properties of finite nuclei as well as nuclear matter. We have extended
this Lagrangian by allowing for the explicit breaking of chiral symmetry
and by including an additional term which permits us to obtain an axial
coupling constant, $g_A$, which is close to the physical value of 1.26 in
vacuum. We have also
adjoined this Lagrangian to a chiral Lagrangian for the $\rho$ and $a_1$
vector mesons \cite{ko} while maintaining reasonable properties for these
particles. This required a renormalization of the pion field. The final
Lagrangian yielded the Goldberger-Treiman relation at finite density, as
well as in the vacuum.

The effect of explicit chiral symmetry breaking on nuclear matter and nuclei
was, for the most part, small so that their properties continued to be
reasonably well described. As in I, the presence of a term
$(\omega_{\mu}\omega^{\mu})^2$ in the Lagrangian improved the
phenomenology. The low energy $\pi N$ scattering parameters were found to be
in good agreement with the data once the effect of the $\Delta$
resonance was removed, since this is not included in our model.
As regards the modifications of the various quantities in going from the
vacuum to equilibrium nuclear matter density, the substantial reduction in
the $\rho$ and $\omega$ masses predicted by QCD sum rules remains puzzling
since the reduction we obtain is very slight. Further it appears that a
more substantial reduction yields poor results for nuclei
\cite{glu3,fs2,fs3}. On the other hand, we find that the axial coupling is
quenched
to a value close to 1, as indicated experimentally, and the increase in the
pion mass suggested by pionic atom data is obtained if the constant in the
symmetry breaking potential $\epsilon_3\bar{N}N$ is zero or very small and
negative. The latter was favored by other data, particularly the
pion-nucleon sigma term. Thus we conclude that our Lagrangian, in which
chiral symmetry is realized in a linear fashion, embodies the essential
physics necessary to describe quite a wide variety of low energy data.

We acknowledge partial support
from the Department of Energy under grants No. DE-FG02-87ER40328
and DE-FG02-94ER40823. A grant for computing time from the Minnesota
Supercomputer Institute is gratefully acknowledged.

\section*{Appendix A: Scalar Field Mass Matrix}

Setting $\sigma=\sigma_0-\bar{\sigma}$ and $\phi=\phi_0-\bar{\phi}$, we
denote the $2\times2$ mass-squared" matrix for the fluctuating parts of the
scalar fields, $\bar{\sigma}$ and  $\bar{\phi}$, by
\begin{equation}
\left(\matrix{2\alpha&-2\beta\cr
-2\beta&2\gamma\cr}\right)\;.
\end{equation}
The matrix elements can be read off from the field equations (\ref{fieldeq}):
\begin{equation}
\alpha=\frac{2B_0\delta+\epsilon_1'-3\epsilon_2'}{2\sigma_0^2}\ ;
\ \beta=\frac{B_0\delta-3\epsilon_2'}{\phi_0\sigma_0}\ ;
\ \gamma=\frac{B_0(2-\delta)+\epsilon_1'-3\epsilon_2'}{\phi_0^2}\:.
\end{equation}
Defining $r=\sqrt{(\gamma-\alpha)^2+4\beta^2}$, the eigenvalues and
eigenvectors are
\begin{eqnarray}
m^2_>&\pj=&\pj\alpha+\gamma+r\quad;\quad\psi_>=a\bar{\sigma}-b\bar{\phi}
\nonumber\\
m^2_<&\pj=&\pj\alpha+\gamma-r\quad;\quad\psi_<=b\bar{\sigma}+a\bar{\phi}\;,
\end{eqnarray}
where $a/b=(\alpha-\gamma+r)/(2\beta)=2\beta/(\gamma-\alpha+r)$ and
$a^2+b^2=1$.

\section*{Appendix B: The Lagrangian for the Vector Fields}

Here we quote, for reference, the Lagrangian of Ko and Rudaz \cite{ko} for
the vector and axial vector mesons, $\rho$ and $a_1$. They define left and
right gauge fields
\begin{equation}
l_{\mu}=\thalf(\vmg{\rho}_{\mu}+\vm{a}_{\mu})\cdot\vmg{\tau}\quad;\quad
r_{\mu}=\thalf(\vmg{\rho}_{\mu}-\vm{a}_{\mu})\cdot\vmg{\tau}\;,
\end{equation}
and field strength tensor
\begin{equation}
l_{\mu\nu}=\partial_{\mu}l_{\nu}-\partial_{\nu}l_{\mu}-if[l_{\mu},
l_{\nu}]\;,
\end{equation}
with an analogous definition for $r_{\mu\nu}$. The Lagrangian is then
\begin{eqnarray}
{\cal L}_{{\rm vector}}&\pj=&\pj-\tquar\:{\rm Tr}[l_{\mu\nu}^2+r_{\mu\nu}^2]
+\thalf m_0^2\:{\rm Tr}[l_{\mu}^2+r_{\mu}^2]\nonumber\\
&&+\tquar bf^2\:{\rm Tr}[\Sigma\Sigma^{\dagger}]\:{\rm Tr}[l_{\mu}^2
+r_{\mu}^2]-cf^2\:{\rm Tr}[l_{\mu}\Sigma r^{\mu}\Sigma^{\dagger}]\nonumber\\
&&-i\frac{\kappa_6f}{4m_{\rho}^2}\:{\rm Tr}
[l_{\mu\nu}\Delta^{\mu}\Sigma\Delta^{\nu}\Sigma^{\dagger}+
r_{\mu\nu}\Delta^{\mu}\Sigma^{\dagger}\Delta^{\nu}\Sigma]\;,\label{lvector}
\end{eqnarray}
using the definitions of Eqs. (\ref{SigPi}) and (\ref{deriv}).
For simplicity we focus here on
the minimal model of Ref. \cite{ko} which neglects a further term in the
Lagrangian ($\zeta_6=0$).
In the present work we have set $b=c$ (see Table 1 for the values)
and generated the dimensionful constants in a scale
invariant way, {\it i.e.} replaced $m_0^2$ by $m_{\rho}^2(\phi/\phi_0)^2$
and $\kappa_6/m_{\rho}^2$ by $(\kappa_6/m_{\rho}^2)(\phi_0/\phi)^2$.

We should briefly comment on the vector meson phenomenology with our chosen
parameters. The predictions for the widths and the pion radius discussed in
Ref. \cite{ko}  are given in Table 4 for our parameters (note that
the corrected formula \cite{eek,song} is used for the
$a_1\rightarrow\rho\pi$ decay  width). In general the results are
reasonable, although the $a_1\rightarrow\pi\gamma$ width is too large by a
factor of two. This width is sensitive to the presence of a $\zeta_6$ term
in the Lagrangian which leads to a renormalization of the $\rho$ and $a_1$
fields. However, inclusion of this term to improve the phenomenology is
beyond the scope of the present work.

\newpage
\begin{center}
{\bf Table 1}\hfill\\
Values of the parameters and derived quantities in nuclear matter.
\end{center}
\begin{center}
\begin{tabular}{|l|c|c|} \hline
Quantity& $G_4=0$ & $G_4/g_{\omega}=0.19$\\ \hline
$|\epsilon_{\rm vac}|^{1/4}$ (MeV)&236&221\\
$g_{\omega}$&10.5&12.4\\
$\zeta=\phi_0/\sigma_0$&1.3&1.4\\
$M^*_{\rm sat}/M$& 0.70& 0.64\\
$K$ (MeV) &388 &317\\
$S/K$  & 6.6&5.4\\
$\sigma_0$ (MeV) &110&102\\
$Z_{\pi}$ &0.71&0.83\\
$f$ & 5.97 &4.95\\
$D$ & 1.36 & 1.10\\
$c$ & 0.57 &1.30\\
$\kappa_6$ &$-1.52$ & $-1.90$ \\ \hline
\end{tabular}
\end{center}
\newpage
\setlength{\oddsidemargin}{-0.8in}
\setlength{\evensidemargin}{-0.8in}
\begin{center}
{\bf Table 2}\hfill\\
Bulk properties of nuclei, sigma term and vacuum axial vector coupling constant
with various explicit symmetry breaking parameters.
\end{center}
\begin{center}
\begin{tabular}{|l|lcr|lr|cc|cc|cc|} \hline
Case&$m_{\pi}$&$\epsilon_2'$&$\epsilon_3$&$g_A$&$\Sigma(0)$
&\multicolumn{2}{|c|}{O}&\multicolumn{2}{c|}{Ca}&\multicolumn{2}{c|}{Pb}\\
&&&&&&$\frac{BE}{A}$&$r_{ch}$&$\frac{BE}{A}$&$r_{ch}$&$\frac{BE}{A}$&$r_{ch}$\\
&(MeV)&&(MeV)&&(MeV)&(MeV)&fm&(MeV)&fm&(MeV)&fm\\ \hline
&Expt.&&&1.26&&7.98&2.73&8.55&3.48&7.86&5.50\\ \hline
A&0 &0 & 0 &1.33& 0 & 7.35 & 2.64 & 7.96 & 3.41 & 7.30 & 5.49\\
F&138 & 0 & $-15$ &1.31& 41 & 7.40 & 2.63 & 7.99 & 3.41 & 7.34 & 5.49\\ \hline
WA&0 &0 & 0 & 1.33& 0 & 7.02& 2.68& 7.75& 3.45& 7.33& 5.51\\
WB&138 & 0 & 0 &1.33 &71& 6.58 & 2.72& 7.43& 3.47& 7.18&5.52\\
WC&138 & $\thalf\epsilon_1'$ & 0 &1.33&101&7.46&2.65&8.10&3.42&7.51&5.51\\
WD&138 & 0 & 50 &1.38&120&6.36&2.73&7.26&3.48&7.06&5.53\\
WE&138 & 0 & $-50$&1.27&24&6.43&2.73&7.31&3.48&7.13&5.52\\
WF&138 & 0 & $-15$ &1.31&56&6.58&2.72&7.45&3.47&7.20&5.52\\
\hline
\end{tabular}
\end{center}

\newpage
\begin{center}
{\bf Table 3}\hfill\\
Low energy pion scattering parameters.
\end{center}
\begin{center}
\begin{tabular}{|l|cccc|cccc|} \hline
&$A_0^{(+)}$ &$m_{\pi}B_0^{(+)}$ &$C_0^{(+)}$ & $D_0^{(+)}$ &
$A_0^{(-)}$ &$m_{\pi}B_0^{(-)}$ &$C_0^{(-)}$ & $D_0^{(-)}$ \\
& $(m_{\pi}^{-1})$ & $(m_{\pi}^{-1})$ & $(m_{\pi}^{-3})$ & $(m_{\pi}^{-3})$ &
$(m_{\pi}^{-1})$ & $(m_{\pi}^{-1})$ & $(m_{\pi}^{-3})$ & $(m_{\pi}^{-3})$ \\
\hline
Expt. & 32.3$\pm$0.4 & $-32.5\pm$0.4 & 2.88$\pm$0.02 &2.65$\pm$0.04&
$-10.7\pm$0.4 & 12.0$\pm$0.4 & $-1.76\pm$0.02 & $-1.13\pm$0.04\\
$\Delta$ \cite{ho}& 7.0  &$-5.4$& 0.98& 0.65& $-10.7$& 10.3 &
$-0.64$& $-0.85$\\
Difference & 25.3 & $-27.1$ & 1.89 & 2.00& 0.1 & 1.7& $-1.11$& $-0.28$\\ \hline
${\cal L}_1$ & 10.7 &$-10.7$ & 1.18 & 0.79 & --& 0.8 & $-0.39$& $-0.12$\\
${\cal L}_1+{\cal L}_2$& 26.4& $-26.4$& 1.76& 1.97& --& 0.8& $-0.97$&
$-0.29$\\
Case A & 26.4 & $-26.4$& 1.64 & 1.97 & -- & 1.1& $-0.97$& $-0.29$\\
Case F& 25.9& $-26.5$& 1.79& 1.97& --& 1.2& $-1.09$& $-0.29$\\ \hline
Case WA&26.4&$-$26.4&1.87&1.97&--&1.1&$-$0.97&$-$0.28\\
Case WB&25.8&$-$26.5&2.01&1.97&--&1.1&$-$1.10&$-$0.29\\
Case WC&26.4&$-$26.5&1.99&1.97&--&1.1&$-$1.10&$-$0.29\\
Case WD&26.6&$-$26.5&2.00&1.97&--&1.0&$-$1.10&$-$0.29\\
Case WE&24.9&$-$26.5&2.06&1.97&--&1.3&$-$1.10&$-$0.29\\
Case WF&25.6&$-$26.5&2.02&1.97&--&1.2&$-$1.10&$-$0.29\\ \hline
\end{tabular}
\end{center}
\newpage
\begin{center}
{\bf Table 4}\hfill\\
Predicitions for the vector mesons with our parameters.
\end{center}
\begin{center}
\begin{tabular}{|l|cc|c|} \hline
Quantity& $G_4=0$ & $G_4/g_{\omega}=0.19$&Experiment\cite{pdg}\\ \hline
$\Gamma(a_1\rightarrow\rho\pi)$ (MeV)&306 &601&$\sim400$\\
$a_1\rightarrow\rho\pi$ $D/S$ ratio&$-0.25$&$-0.17$&$-0.14\pm0.03$
\cite{dtos}\\
$\Gamma(a_1\rightarrow\pi\gamma)$ (MeV)& 1.10 & 1.19 &$0.64\pm0.25$\\
$\Gamma(\rho^0\rightarrow e^+e^-)$ (keV) &4.82 & 7.00 & $6.77\pm0.32$ \\
$\langle r^2\rangle^{\frac{1}{2}}_{\stackrel{~}{\pi}}$ (fm) & 0.63 & 0.69 &
$0.66\pm0.01$\\ \hline
\end{tabular}
\end{center}
\vskip1in

\centerline{{\bf Figure Captions}}

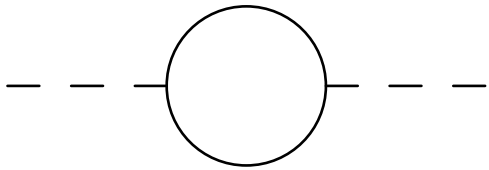
\noindent Figure 1. Nucleon loop contribution to the pion propagator.

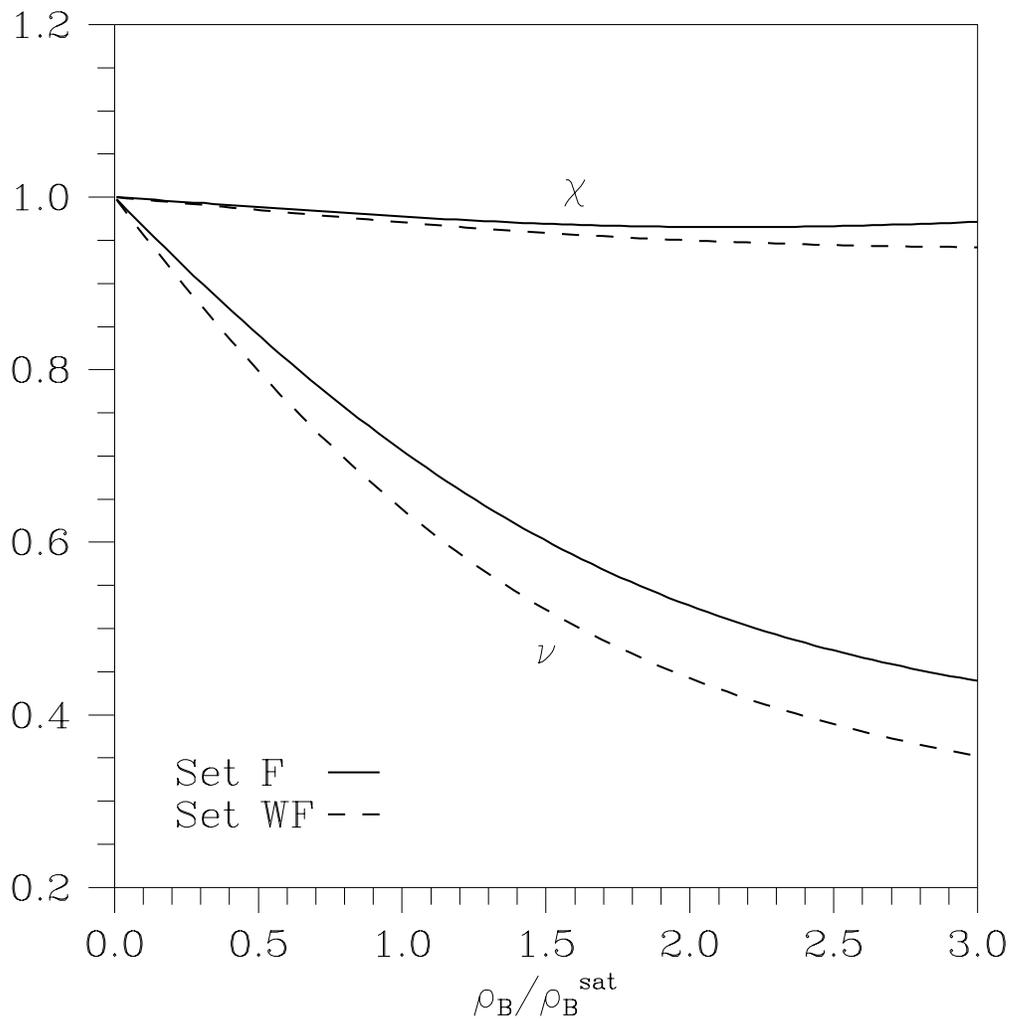
\noindent Figure 2. Ratio of glueball and sigma fields to their vacuum
values, $\chi$ and $\nu$, as a function of density. The parameters F and WF
are specified in the text. Equilibrium nuclear matter density is denoted by
$\rho_B^{\rm sat}$.

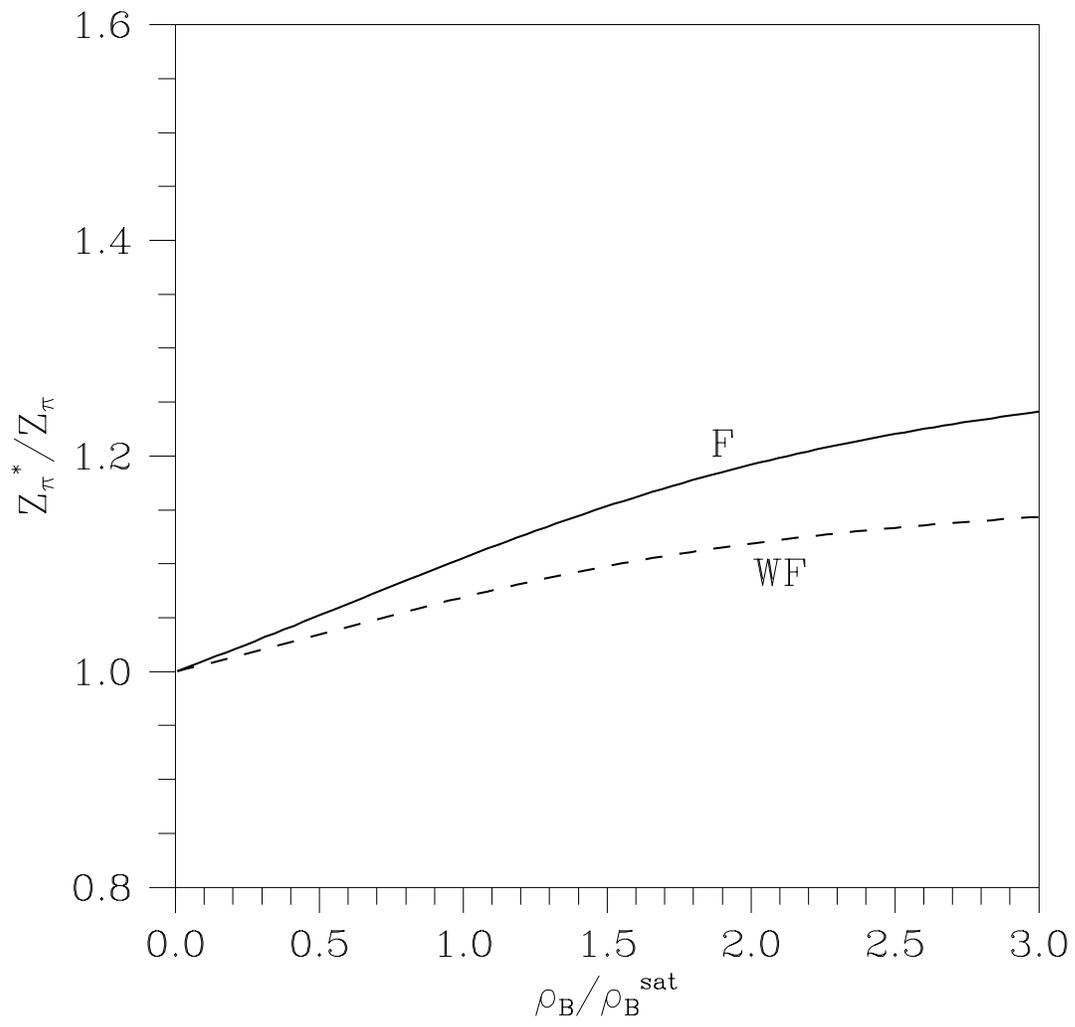
\noindent Figure 3. Ratio of the pion renormalization $Z_{\pi}^*$ in matter to
the vacuum value $Z_{\pi}$ as a function of density.

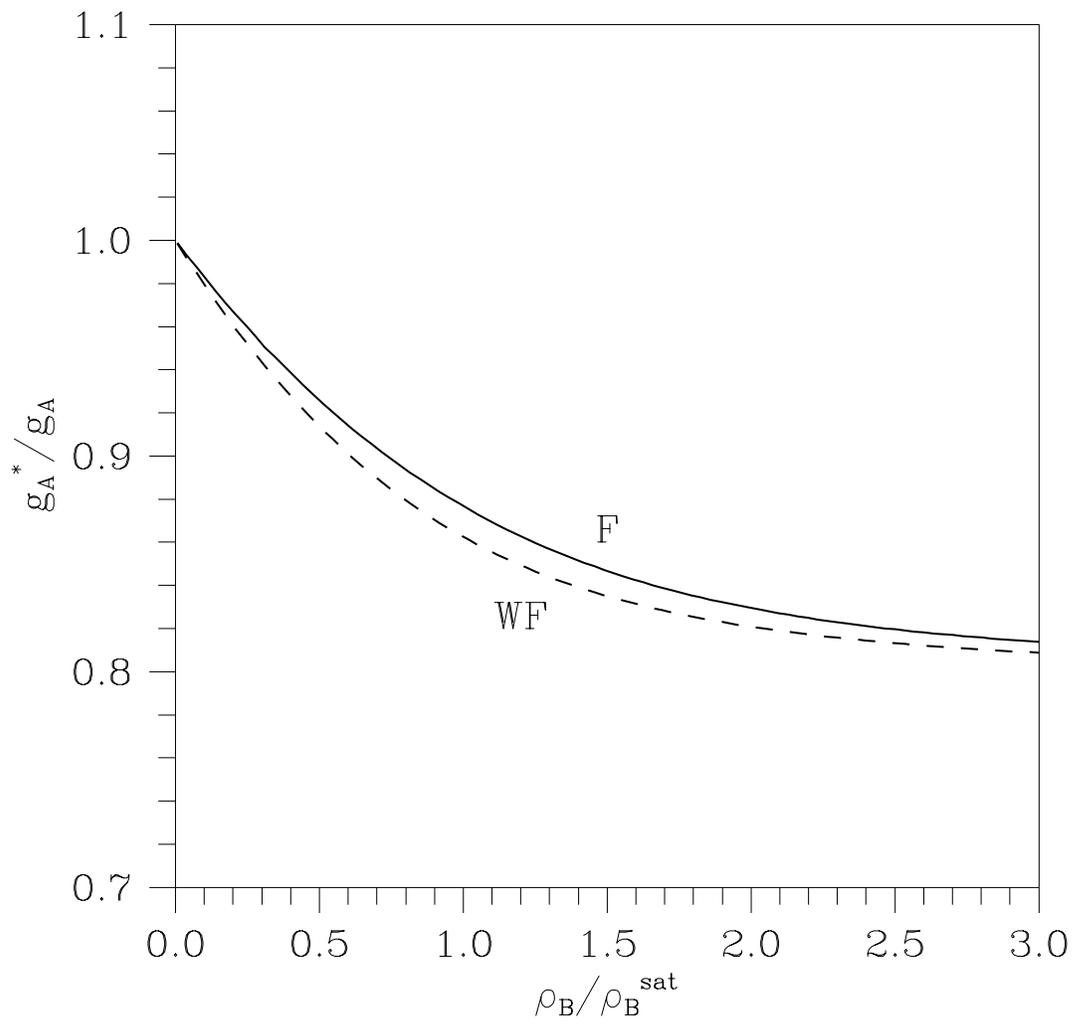
\noindent Figure 4. Ratio of the axial coupling constant $g_A^*$ in matter to
the vacuum value $g_A$ as a function of density.

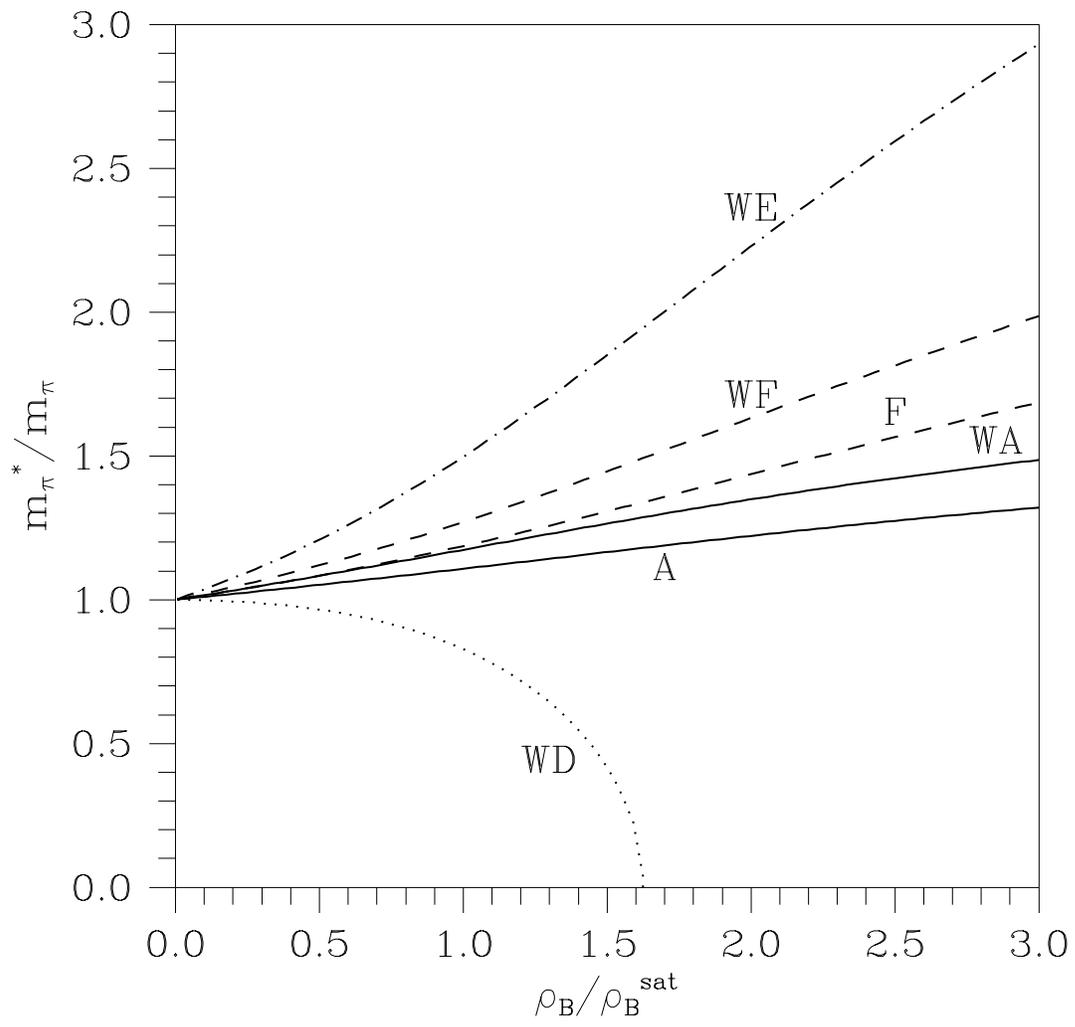
\noindent Figure 5. Ratio of the pion mass $m_{\pi}^*$ in matter to
the vacuum value $m_{\pi}$ as a function of density. The parameter sets are
specified in the text.


\begin{thebibliography}{999}
\bibitem{kerman} A.K. Kerman and L.D. Miller, ``Field Theory Methods for
Finite Nuclear Systems and the Possibility of Density Isomerism", in {\it
Proceedings of the Relativistic Heavy Ion Summer Study}, Berkeley report
LBL--3675 (1974) p.73.
\bibitem{glu3} E.K. Heide, S. Rudaz and P.J. Ellis,
Nucl. Phys. {\bf A571} (1994) 713.
\bibitem{fs2} R.J. Furnstahl and B.D. Serot,
Phys. Rev. {\bf C47} (1993) 2338; Phys. Lett. {\bf B316} (1993) 12.
\bibitem{fs3} R.J. Furnstahl, B.D. Serot and H.-B. Tang,
preprint (1995) nucl-th/9511028.
\bibitem{fern2} R.J. Furnstahl, C.E. Price and G.E. Walker,
Phys. Rev. {\bf C36} (1987) 2590.
\bibitem{brook} L.S. Celenza, C.M. Shakin, W.-D. Sun and J. Szweda,
Brooklyn College of the City University of New York preprint,
\#BCCNT 95/032/245; C.M. Shakin, W.-D. Sun and J. Szweda, {\it ibid.}
\#BCCNT 95/051/246.
\bibitem{fst} R.J Furnstahl, H.-B. Tang and B.D. Serot,
Phys. Rev. {\bf C52} (1995) 1368 and work in progress.
\bibitem{scalar} M. Svec, preprint (1995) hep-ph/9511205;
N.A. Tornqvist and M. Roos, preprint (1995) hep-ph/9511210.
\bibitem{lee} B.W. Lee, {\it Chiral Dynamics} (Gordon and Breach, NY, 1972)
\bibitem{gas} S. Gasiorowicz and D.A. Geffen,
Rev. Mod. Phys. {\bf41} (1969) 531.
\bibitem{velo} G. Velo and D. Zwanziger, Phys. Rev. {\bf188} (1969) 2218;
G. Velo, Nucl. Phys. {\bf B65} (1973) 427.
\bibitem{glu2} E.K. Heide, S. Rudaz and P.J. Ellis,
Phys. Lett. {\bf B293} (1992) 259.
\bibitem{mig} J. Schechter, Phys. Rev. {\bf D21} (1980) 3393; A.A. Migdal and
M.A. Shifman, Phys. Lett. {\bf B114} (1982) 445.
\bibitem{gomm} H. Gomm and J. Schechter, Phys. Lett. {\bf B158} (1985) 449.
\bibitem{horse} C.J. Horowitz and B.D. Serot, Nucl. Phys. {\bf A368} (1981)
503.
\bibitem{sew} B.D. Serot and J.D. Walecka, Advances in Nuclear Physics,
Vol. 16, ed. J.W. Negele and E. Vogt (Plenum, NY, 1986); B.D. Serot,
Rep. Prog. Phys. {\bf 55} (1992) 1855.
\bibitem{gold} M. Goldberger and S.B. Treiman,
Phys. Rev. {\bf110} (1958) 1178.
\bibitem{ak} E.Kh. Akhmedov, Nucl. Phys. {\bf A500} (1989) 596.
\bibitem{ko} P. Ko and S. Rudaz,
Phys. Rev. {\bf D50} (1994) 6877.
\bibitem{mat} T. Matsui and B.D. Serot, Ann. Phys. (NY) {\bf144} (1982) 107.
\bibitem{cam} D.K. Campbell, Phys. Rev. {\bf C19} (1979) 1965.
\bibitem{svz} M.A. Shifman, A.I. Vainshtein and V.I. Zakharov, Nucl. Phys.
{\bf B147} (1979) 385, 448.
\bibitem{bonn} R. Machleidt, K. Holinde and Ch. Elster,
Phys. Rep. {\bf149} (1987) 1.
\bibitem{shif} M.A. Shifman, Z. Phys. {\bf C9} (1981) 347; P. Pascual and
R. Tarrach, Phys. Lett. {\bf B113} (1982) 495.
\bibitem{lein} D.B. Leinweber, X. Jin and R.J. Furnstahl, preprint
(1995) nucl-th/9511031.
\bibitem{comp} J.P. Blaizot, Phys. Rep. {\bf64} (1980) 171;
N.K. Glendenning, Phys. Rev {\bf C37} (1988) 2733;
M.V. Stoitsov, P. Ring and M.M. Sharma, Phys. Rev. {\bf C50} (1994) 1445.
\bibitem{pear} J.M. Pearson, Phys. Lett. {\bf B271} (1991) 12;
see also S. Rudaz, P.J. Ellis, E.K. Heide and M. Prakash, Phys. Lett.
{\bf B285} (1992) 183.
\bibitem{dum} O. Dumbrajs, R. Koch, H. Pilkuhn, G.C. Oades, H. Behrens,
J.J. de Swart and P. Kroll, Nucl. Phys. {\bf B216} (1983) 277.
\bibitem{neg} J.W. Negele, Phys. Rev. {\bf C1} (1970) 1260;
L.J. Tassie and F.C. Barker, Phys. Rev. {\bf111} (1958) 940.
\bibitem{ho} G. H\"ohler, {\it Pion-Nucleon Scattering}, Landolt-B\"ornstein,
New Series, ed. H. Schopper, Vol. I/9 b2 (Springer-Verlag, Berlin, 1983).
\bibitem{gass} J. Gasser, H. Leutwyler and M.E. Sainto,
Phys. Lett. {\bf B253} (1991) 252.
\bibitem{pdg} Particle Data Group, L. Montanet {\it et al.}, Phys. Rev.
Phys. Rev. {\bf D50} (1994) 1173.
\bibitem{qcdvec} T. Hatsuda and S.H. Lee, Phys. Rev. {\bf C46} (1992) R34;
X. Jin and D.B. Leinweber, preprint (1995) nucl-th/9510064.
\bibitem{scaling} G.E. Brown and M. Rho,
Phys. Rev. Lett. {\bf66} (1991) 2720.
\bibitem{price} C.E. Price, J.R. Shepard and J.A. McNeil,
Phys. Rev. {\bf C41} (1990) 1234; {\bf C42} (1990) 247.
\bibitem {quench} B. Buck and S.M. Perez, Phys. Rev. Lett. {\bf50} (1983)
1975; M. Rho, Ann. Rev. Nucl. Sci. {\bf34} (1984) 531.
\bibitem{vest} V. Thorsson and A. Wirzba,
Nucl. Phys. {\bf A589} (1995) 633.
\bibitem{eric} J. Delorme, M. Ericson and T.E.O. Ericson,
Phys. Lett. {\bf B291} (1992) 379.
\bibitem{eek} V.L. Eletsky, P.J. Ellis and J.I. Kapusta,
Phys. Rev. {\bf D47} (1993) 4084.
\bibitem{song} C. Song, Phys. Rev. {\bf C47} (1993) 2861.
\bibitem{dtos} N. Isgur, C. Morningstar and C. Reader,
Phys. Rev. {\bf D39} (1989) 1357.
\end{thebibliography}
\end{document}